\documentclass{aa}
\usepackage{graphicx}
\usepackage{psfrag}
\usepackage{txfonts}
\usepackage{natbib}
\bibpunct{(}{)}{;}{a}{}{,} 
\usepackage{url}
\begin{document}
    \title{Limits of ultra-high-precision optical astrometry:\\
    Stellar surface structures}

    \author{U. Eriksson
          \inst{1}\, \inst{2}
          \and
          L. Lindegren
          \inst{1}
          }

   \offprints{U. Eriksson, \email{urban@astro.lu.se}}

     \institute{Lund Observatory, Lund University, Box 43, SE-221 00 Lund, Sweden\\
         \and
             Dep. of Mathematics and Science, Kristianstad University, SE-291 88 Kristianstad, Sweden\\
             }

   \date{Received Month day, 2007; accepted Month day, 2007}


  \abstract
   {}
   {
    To investigate the astrometric effects of stellar surface
    structures as a practical limitation to ultra-high-precision
    astrometry, e.g.\ in the context of exoplanet searches, and to
    quantify the expected effects in different regions of the
    HR-diagram. }
   { Stellar surface structures (spots, plages, granulation,
   non-radial oscillations) are likely to produce fluctuations in
   the integrated flux and radial velocity of the star, as well as
   a variation of the observed photocentre, i.e.\ astrometric jitter.
   We use
   theoretical considerations supported by Monte Carlo simulations
   (using a starspot model) to derive statistical
   relations between the corresponding astrometric, photometric,
   and radial-velocity effects. Based on these relations, the more
   easily observed photometric and radial-velocity variations can
   be used to predict the expected size of the astrometric jitter.
   Also the third moment of the brightness distribution,
   interferometrically observable as closure phase, contains
   information about the astrometric jitter.}
   {For most stellar types the astrometric jitter due to stellar surface
   structures is expected to be of order 10~micro-AU or greater. This is
   more than the astrometric displacement typically caused by an
   Earth-size exoplanet in the habitable zone, which is about
   1--4~micro-AU for long-lived main-sequence stars. Only
   for stars with extremely low photometric variability
   ($<0.5$~mmag) and low magnetic activity, comparable to that of
   the Sun, will the astrometric jitter be of order 1~micro-AU,
   sufficient to allow the astrometric detection of an Earth-sized
   planet in the habitable zone. While stellar surface structure
   may thus seriously impair the astrometric detection of small
   exoplanets, it has in general negligible impact on the detection
   of large (Jupiter-size) planets and on the determination of
   stellar parallax and proper motion. From the starspot model we
   also conclude that the commonly used spot filling factor is not
   the most relevant parameter for quantifying the spottiness in
   terms of the resulting astrometric, photometric and radial-velocity
   variations.
   }
   {}


   \maketitle
%

\section{Introduction}

The accuracy of astrometric measurements has improved tremendously in the past decades as a result of new techniques being
introduced, both on the ground and in space. This development will continue in the next decade, e.g\ Gaia is to improve
parallax accuracy by another two orders of magnitude compared with Hipparcos. As a result, trigonometric distances will be
obtained for the Magellanic Clouds, and thousands of Jupiter-size exoplanets are likely to be found from the astrometric
wobbles of their parent stars. Even before that, ground-based interferometric techniques are expected to reach similar
precisions for relative measurements within a small field. How far should we expect this trend to continue? Will nanoarcsec
astrometry soon be a reality, with parallaxes measured to cosmological distances and Earth-size planets found wherever we
look? Or will the accuracy ultimately be limited by other factors such as variable optical structure in the targets and weak
microlensing in the Galactic halo? The aim of this project is to assess the importance of such limitations for
ultra-high-precision astrometry. In this paper we consider the effects of stellar surface structures found on ordinary stars.

Future high-precision astrometric observations will in many cases be able to detect the very small shifts in stellar
positions caused by surface structures. In some cases, e.g.\ for a rotating spotted star, the shifts are periodic and could
mimic the dynamical pull of a planetary companion, or even the star's parallax motion, if the period is close to one year.
These shifts are currently of great interest as a possible limitation of the astrometric method in search for Earth-like
exoplanets. We want to estimate how important these effects are for different types of stars, especially in view of current
and future astrometric exoplanet searches such as VLTI-PRIMA \citep{Reffert_2005ASPC..338...81R}, SIM PlanetQuest
\citep{Unwin_2005ASPC..338...37U} and Gaia \citep{Lattanzi_2005}.

Astrometric observations determine the position of the centre of gravity of the stellar light, or what we call the
photocentre. This is an integrated property of the star (the first moment of the intensity distribution across the disk), in
the same sense as the total flux (the zeroth moment of the intensity distribution) or stellar spectrum (the zeroth moment as
function of wavelength). In stars other than the Sun, information about surface structures usually come from integrated
properties such as light curves and spectrum variations. For example, Doppler imaging (DI) has become an established
technique to map the surfaces of rapidly rotating, cool stars. Unfortunately, it cannot be applied to most of the targets of
interest for exoplanet searches, e.g.\ low-activity solar-type stars. Optical or infrared interferometric (aperture
synthesis) imaging does not have this limitation, but is with current baselines ($<1$~km) in practice limited to giant stars
and other extended objects (see Monnier et al.\ 2006 for a review on recent advances in stellar interferometry).
Interferometry of marginally resolved stars may, however, provide some information about surface structures through the
closure phase, which is sensitive to the third central moment (asymmetry) of the stellar intensity distribution
\citep{Monnier2003RPPh...66..789M,Lachaume2003A&A...400..795L, Labeyrie2006iai..book.....L}.

Since there is limited information about surface structures on most types
of stars, an interesting question is whether we can use more readily accessible
photometric and spectroscopic data to infer something about possible
astrometric effects. For example, dark or bright spots on a rotating star
will in general cause periodic variations both in the integrated flux and in
the radial velocity of the star, as well as in the photocentre and the
asymmetry of the intensity distribution. Thus, we should at least expect
the astrometric effect to be statistically related to the other effects.

We show that there are in fact relatively well-defined statistical relations
between variations in the photocentre, total flux, closure phase and radial
velocity for a wide range of possible surface phenomena. These relations are
in the following used to predict the astrometric jitter in various types of
stars, without any detailed knowledge of their actual surface structures.

\section{Astrometric limits from previous studies}

The discovery of exoplanets by means of high-precision radial velocity measurements has triggered an interest in how
astrophysical phenomena such as magnetic activity and convective motions might affect the observed velocities
\citep{Saar2003csss...12..694S}. Evidence for dark spots have been seen photometrically and spectroscopically for many cool
stars other than the Sun, and quantified in terms of an empirically determined \emph{spot filling factor}%
\footnote{$f$ is interpreted as the fraction of the visible hemisphere of the star covered by spots.}%
$f$, ranging from $\ll 1\%$ for old, inactive stars to several percent for active stars. It is therefore natural to relate
the expected radial-velocity effects to the spot filling factor. For example, \citet{Saar_Donahue_1997} used a simple model
consisting of a single black equatorial spot on a rotating solar-like star to derive the following relation between $f$ (in
percent), the projected rotational velocity $V\sin i$ and the amplitude $\Delta v_r$ of the resulting radial velocity
variations:
\begin{equation}
\Delta v_r=0.0065\,f^{0.9}\,V\sin i
\label{Saar 1}
\end{equation}
In a similar vein, \citet{Hatzes_2002AN....323..392H} estimated both
the radial velocity amplitude and the corresponding astrometric effect
from a similar model, but assuming a fixed spot size ($2^\circ$ radius)
and instead varying the number of spots placed randomly on the stellar
surface centred around the equator. For the radial velocity amplitude
they found
\begin{equation}
\Delta v_r=\left(0.0086\,V\sin i-0.0016\right)\,f^{0.9}
\label{Hatzes 1}
\end{equation}
in approximate agreement with (\ref{Saar 1}), while the total amplitude
of the astrometric effect (converted to linear distance) was
\begin{equation}
\Delta{\rm pos}=(7.1\times 10^{-5}~\mbox{AU})\,f^{0.92}
\label{Hatzes 2}
\end{equation}

\citet{Reffert_2005ASPC..338...81R} discuss the accuracy and limitations of the
PRIMA (Phase-Referenced Imaging and Micro-Arcsecond Astrometry) facility at the
VLT Interferometer in the context of the search for suitable targets for
exoplanetary searches, reference and calibrations stars. According to their
calculations, a spot filling factor of $f=2$\% would move the photocentre of
a G0V star by about $3\times 10^{-5}$~AU, roughly a factor 4 less than
according to (\ref{Hatzes 2}). They also conclude that the corresponding
brightness variation is less than 2\%.

But $f$ alone may not be a very good way to quantify the `spottiness'.
For example, the photometric or astrometric effects of a large single spot
are obviously very different from those of a surface scattered with many
small spots, although the spot filling factor may be the same in the two
cases. Therefore, more detailed (or more general) models may be required
to explore the plausible ranges of the astrometric effects.

\citet{Bastian2005} give an assessment of the astrometric effects of starspots, and conclude that they are hard to quantify,
mostly because of the insufficient statistics. Although starspots are common among cool stars with outer convective zones,
data are strongly biased towards very active stars. They conclude that the effects on solar-type stars are likely to be
negligible for Gaia, while much larger spots on K giants may become detectable. For supergiants and M giants, having radii of
the order of $100R_\odot$ (or more), the effect may reach 0.25~AU (or more), which could confuse the measurement of parallax
and proper motion.

\citet{sozzetti2005} gives an interesting review of the astrometric
methods to identify and characterize extrasolar planets. As an example
of the astrophysical noise sources affecting the astrometric measurements,
he considers a distribution of spots on the surface of a pre-mainsequence
(T~Tauri) star. For a star with radius $1R_{\odot}$ seen at a distance of
140~pc, he finds that a variation of the flux in the visual by
$\Delta F/F=10$\% (rms) corresponds to an astrometric variation of
$\sim\! 3~\mu$as (rms), and that the two effects are roughly proportional.

While the astrometric effects cannot yet be tested observationally, it is possible to correlate the photometric and
radial-velocity variations for some stars \citep{Queloz2001A&A...379..279Q, Henry2002ApJ...577L.111H}. From a small sample of
Hyades stars \citet{Paulson_Saar2004AJ....127.1644P} found an approximately linear relation
\begin{equation}
\sigma_{v_{\rm R}} \simeq 2 + 3600\,\sigma_m  \quad \mbox{[m~s$^{-1}$]}
\label{Paulson fig4}
\end{equation}
between the RMS scatter in Str{\"o}mgren $y$ magnitude ($\sigma_m$) and in
radial velocity ($\sigma_{v_{\rm R}}$). This relation
supports the idea that a large part of the radial-velocity scatter
in these stars is caused by surface structures.

\citet{Svensson_Ludwig_2005} have computed hydrodynamical model atmospheres for a range of stellar types, predicting both the
photometric and astrometric jitter caused by granulation. They find that the computed astrometric jitter is almost entirely
determined by the surface gravity $g$ of the atmosphere model, and is proportional to $g^{-1}$ for a wide range of models.
This relationship is explained by the increased granular cell size with increasing pressure scale height or decreasing $g$.
The radius of the star does not enter the relation, except via $g$, since the increased leverage of a large stellar disk is
compensated by the averaging over more granulation cells. For their most extreme model, a bright red giant with $\log g=1$
($R/R_\odot\simeq 95$) they find $\sigma_{\rm pos}\simeq 300~\mu$AU. \citet{Ludwig2005_2} extended this by considering the
effects of granulation on interferometric observations of red supergiants. They show that both visibilities and closure
phases may carry clear signatures of deviations from circular symmetry for this type of stars, and conclude that
convection-related surface structures may thus be observable using interferometry.

\citet{Ludwig2005} outlines a statistical procedure to characterise
the photometric and astrometric effects of granulation-related
micro-variability in hydrodynamical simulations of convective stars.
Based on statistical assumptions similar to our model in Appendix~A,
he finds the relation
\begin{equation}
\frac{\sigma_x}{R} \simeq \frac{1}{\sqrt{6}}
\frac{\sigma_{F}}{\left\langle F\right\rangle}  \label{H.-Gs relation}
\end{equation}
between the RMS fluctuation of the photocentre in one coordinate ($x$),
the radius of the star ($R$), and the relative fluctuations of the
observed flux ($F$).

\section{Modeling astrometric displacements}\label{theory}

\subsection{Relations for the astrometric jitter}\label{general relations}

In a coordinate system $\vec{xyz}$ with origin at the centre of the star and $+\vec{z}$
away from the observer, let $I(\vec{r},t)$ be the instantaneous surface brightness
of the star at point $\vec{r}=(x,y,z)$ on the visible surface, i.e.\ the specific
intensity in the direction of the observer. We are interested in the integrated
properties: total flux $F(t)$, photocentre offsets $\Delta x(t)$, $\Delta y(t)$ in
the directions perpendicular to the line of sight, the third central moment of the
intensity distribution $\mu_3(t)$, and the radial velocity offset $\Delta v_{\rm R}(t)$.
These are given by the following integrals over the visible surface $S$ ($z<0$):
\begin{eqnarray}
F(t)   &=& \int_S I(\vec{r},t)\mu\,\mbox{d}S \label{Intensity1} \\
\Delta x(t) &=&\frac{1}{F(t)} \int_S I(\vec{r},t)x\mu\,\mbox{d}S \label{disp_x_1} \\
\Delta y(t) &=&\frac{1}{F(t)} \int_S I(\vec{r},t)y\mu\,\mbox{d}S \label{disp_y_1}\\
\mu_3(t)  &=&\frac{1}{F(t)} \int_S I(\vec{r},t)\left[ x-\Delta x(t)\right]^3\mu\,
\mbox{d}S \label{disp_x_3}\\
\Delta v_{\rm R}(t)&=&\frac{1}{F(t)} \int_S I(\vec{r},t)
\left[(\vec{\omega}\times\vec{r})\cdot\hat{\vec{z}}\right]\,\mu\,\mbox{d}S\label{RV1}
\end{eqnarray}
where $\mu=|z|/R$ is the geometrical projection factor applied to the surface element
when projected onto the sky, $\vec{\omega}$ is the angular velocity of the star and
$\hat{\vec{z}}$ the unit vector along $+\vec{z}$. (For the third moment, only the
pure $x$ component is considered above.) Equation (\ref{RV1}) assumes that
the star rotates as a rigid body, that rotation is the only cause of the
radial-velocity offset, and that the overall offset can be calculated as the
intensity-weighted mean value of the local offset across the surface. The flux
variation expressed in magnitudes is
\begin{equation}
\Delta m(t)=1.086\frac{F(t)-\left\langle F\right\rangle }{\left\langle F\right\rangle }
\end{equation}
where $\left\langle F\right\rangle $ is the time-averaged flux.
\nocite{Gray2005oasp.book.....G}

Using a similar statistical method as \citet{Ludwig2005}, the RMS variations
(dispersions) of $m(t)$, $\Delta x(t)$, $\Delta y(t)$ and $\mu_3(t)$ can be
estimated from fairly general assumptions about the surface brightness
fluctuations (Appendix \ref{appendixA}). This calculation is approximately valid
whether the fluctuations are caused by dark or bright spots, granulation, or a
combination of all three, and whether or not the time variation is caused by the
rotation of the star or by the changing brightness distribution over the surface.
The result is a set of proportionality relations involving the radius of the star
$R$, the limb-darkening factor $a$, and the centre-to-limb variation $c$ of the
surface structure contrast [see (\ref{a5}) and (\ref{a18}) for the definition
of $a$ and $c$]. For $a=0.6$ (typical solar limb-darkening in visible light) and
$c=0$ (no centre-to-limb variation of contrast) we find
\begin{eqnarray}
\sigma _{\Delta x}=\sigma _{\Delta y} \equiv \sigma_{\mathrm{pos}} &\simeq&
0.376\,R\,\sigma_m \label{theor_pos}\\
\sigma _{\mu_3} &\simeq& 0.139\,R^3\,\sigma_m \label{theor_3rd}
\end{eqnarray}
where $\sigma_q$ designates the dispersion of the quantity $q$.

For the radial-velocity dispersion, a similar relation can be derived under the
previously mentioned conditions of a time-independent, rigidly rotating star.
Using that $(\vec{\omega}\times\vec{r})\cdot\hat{\vec{z}}=\omega_x y - \omega_y x$
we have
\begin{equation}\label{eq:vr_xy}
\Delta v_{\rm R}(t) = \omega_x \Delta y(t) - \omega_y \Delta x(t)
\end{equation}
and
\begin{equation}\label{eq15}
\sigma_{v_{\rm R}}^2 = \omega_x^2 \sigma_y^2 + \omega_y^2 \sigma_x^2 =
(\omega_x^2 + \omega_y^2) \sigma_{\rm pos}^2
\end{equation}
since $\Delta x(t)$ and $\Delta y(t)$ are statistically uncorrelated according
to Eq.~(\ref{a7}).
Noting that $R(\omega_x^2 + \omega_y^2)^{1/2}$ equals
the projected rotational velocity $V\sin i$ we can also write
(\ref{eq15}) as
\begin{equation}\label{eq16}
\sigma_{\rm pos}=R\sigma _{v_{\rm R}}/(V\sin i)
\end{equation}
which may be used to predict the astrometric jitter from the radial velocity variations, if the latter are mainly caused by
rotational modulation. Combined with (\ref{theor_pos}) we find under the same assumption
\begin{equation}\label{vr_vsini}
\sigma _{v_{\rm R}}\simeq 0.376 V\sin i \,\sigma_m
\end{equation}
In terms of the rotation period $P=2\pi/\omega$, and assuming random
orientation of $\vec{\omega}$ in space, Eq.~(\ref{eq16}) can be written
\begin{equation}
\sigma _{\mathrm{pos}}=\sqrt{\frac{3}{2}}\;\frac{P}{2\pi}\,\sigma_{v_{\rm R}}
\simeq 0.195\,P\,\sigma_{v_{\rm R}}
 \label{theor_vr}
\end{equation}

\subsection{Modeling discrete spots}\label{model}

\begin{figure}
\resizebox{\hsize}{!}{
\includegraphics[]{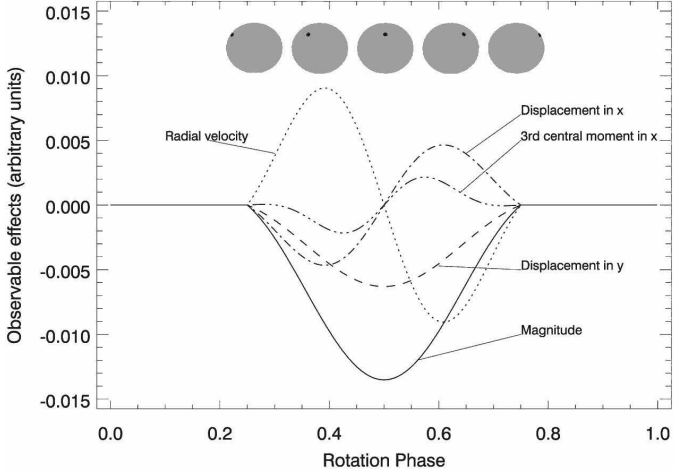}}
\caption{The curves show the effects in magnitude, position, radial velocity and intensity skewness (third central moment) of
a single dark spot located at latitude $30^\circ$. The star is observed at inclination $i=90^\circ$ and the limb-darkening
parameter $a=0.6$. The vertical scale is in arbitrary units for the different effects.} \label{fig1}
\end{figure}

As a check of the general relations in Sect.~\ref{general relations} we have
made numerical simulations with a very simple model, consisting of a limited
number of (dark or bright) spots on the surface of a rotating star. The
behaviour of the integrated properties are readily understood in this case
(cf.\ Fig.~\ref{fig1}):
\begin{itemize}
\item
the flux is reduced in proportion to the total projected area
of the visible spots (or the spot filling factor $f$);
\item
a black spot on, say, the $+x$ side of the star will shift the photocentre
in the $-x$ direction and cause a negative skewness of the flux distribution
along the $x$ direction;
\item
the apparent radial velocity of the star is modified, depending on whether
the dark spot is located on the part of the disk moving towards the observer
(giving $\Delta v_{\rm R}>0$) or away from the observer ($\Delta v_{\rm R}<0$)
\citep[p.~496 and references therein]{Gray2005oasp.book.....G}.
\end{itemize}
Bright spots cause similar effects but with the opposite sign. Limb darkening
of the stellar disk and a possible centre-to-limb variation of spot contrast
will modify the precise amount of these shifts, but not their qualitative
behaviour.

We assume a spherical star with $N$ spots that are:
\begin{itemize}
\item absolutely black,
\item small compared to the stellar radius $R$,
\item of equal area $A$ (measured as a fraction of the total surface),
\item randomly spread over the whole stellar surface, and
\item fixed in position on the surface, while the star rotates.
\end{itemize}
For circular spots of angular radius $\rho$ (as seen from
the centre of the star), we have $A=\sin^2(\rho/2)$.
The assumption of absolutely black spots is uncritical if we interpret
$A$ as the \emph{equivalent area} of the spot, i.e.\ the area of a completely
black spot causing the same drop in flux. Bright spots can formally be
handled by allowing negative $A$.

The star is assumed to rotate as a rigid body with period $P$ around an axis
that is tilted an angle $i$ to the line of sight ($+z$).
For the present experiments we take the $+y$ direction to coincide with the
projection of the rotation vector $\vec{\omega}$ onto the sky; thus
$\omega_x=0$, $\omega_y=\omega\sin i$, and $\omega_z=\omega\cos i$, where
$\omega=2\pi/P$. Limb darkening of the form intensity $\propto 1-a+a\mu$
is assumed, where $\mu=|z|/R$.

To model a rotating spotted star, we place the $N$ spots of the given size
$A$ randomly on the surface of a spherical star and tilt the axis to a certain
inclination $i$. Letting the star rotate around its axis we calculate the
integrated quantities as functions of the rotational phase, taking into
account the projection effect on the area of each spot (by the factor $\mu$)
as well the limb-darkening law.

The effects of a single black spot as function of the rotational phase are illustrated in Fig.~\ref{fig1}. It can be noted
that the effects are not unrelated to each other; for example, the radial-velocity curve mirrors the displacement in $x$, and
both of these curves look like the derivative of the photometric curve. This is not a coincidence but can be understood from
fairly general relations like (\ref{eq:vr_xy}). With many spots the curves become quite complicated, but some of the basic
relationships between them remain.

\begin{figure}
\resizebox{\hsize}{!}{
\includegraphics[angle=-90]{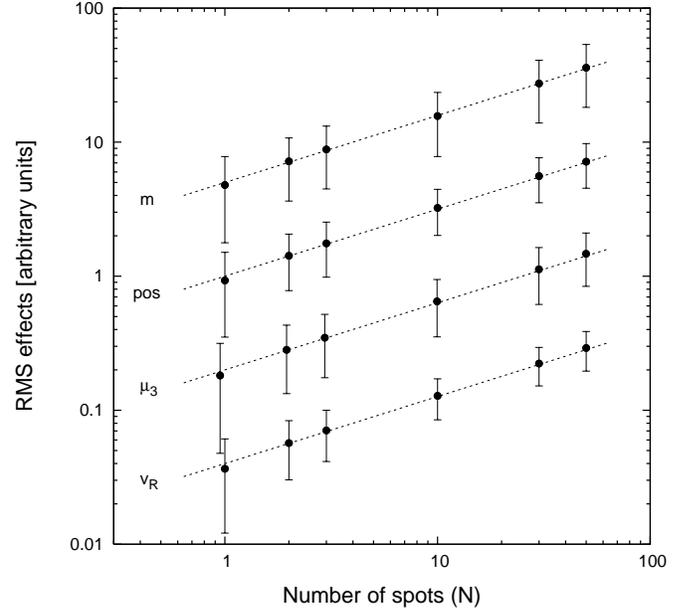}}
\caption{Results of Monte Carlo simulations of rotating stars with different
number ($N$) of spots, all of the same size ($A=0.0025$). The different graphs
refer to (from top to bottom) $\sigma_m$, $\sigma_{\rm pos}$, $\sigma_{\mu_3}$
and $\sigma_{v_{\rm R}}$, expressed on an arbitrary scale; the dots and error
bars show the mean value and dispersion of the $\sigma$ values for a set of
simulations with given $N$. The dashed lines have slope 0.5, corresponding to
$\sigma\propto\sqrt{N}$.}
\label{fig2}
\end{figure}

The total equivalent area of the spots is $AN$ (the spot filling factor $f\simeq 2AN$). As long as $AN \ll 1$, all the
effects are proportional to $A$. The dependence on $N$ is more complex because of the random distribution of spots. For
example, the photometric effect will mainly depend on the actual number of spots $k$ visible at any time. For any random
realization of the model, $k$ follows a binomial distribution with parameters $p=0.5$ and $N$; its dispersion is therefore
$\sqrt{N}/2$. We can therefore expect the RMS photometric effect to be roughly proportional to $A\sqrt{N}$. Similar arguments
(with the same result) can be made for the other effects.

Monte Carlo simulations of a large number of cases with $A=0.0025$
(spot radius $\rho=5.73^\circ$) and $N$ in the range from 1 to 50 (assuming
random orientation of the rotation axis and a limb-darkening parameter
$a=0.6$) indeed show that the RMS effects in magnitude,
photocentre displacements, third central moment and radial
velocity are all, in a statistical sense, proportional to
$\sqrt{N}$ (Fig.~\ref{fig2}). More precisely we find
\begin{eqnarray}
\sigma_m &\simeq& (1.17 \pm 0.60) \cdot A\sqrt{N} \label{res1}\\
\sigma_{\mathrm{pos}} &\simeq& (0.57 \pm 0.25) \cdot A\sqrt{N}\cdot R \label{res2}\\
\sigma_{\mu_3} &\simeq& (0.22 \pm 0.09) \cdot A\sqrt{N}\cdot R^3 \label{res3}\\
\sigma _{v_{\rm R}} &\simeq& (0.51 \pm 0.26) \cdot A\sqrt{N} \cdot R\,\omega \label{res4}  
\end{eqnarray}
where the values after $\pm$ show the RMS dispersion of the
proportionality factor found among the different simulations.

The relations (\ref{res1})--(\ref{res4}) suggest that a measurement
of any one of the four dispersions can be used to \emph{statistically}
predict the other three dispersions, assuming that we know the
approximate radius and rotation period of the star, and that the
different effects are indeed caused by the rotating spotted surface.
An important point is that it is not necessary to know $A$ or $N$
in order to do this. For example, expressing the other effects in
terms of the photometric variation we find
\begin{eqnarray}
\sigma_{\rm pos} &\simeq& 0.49\,R\,\sigma_m \label{theor_pos_sim}\\
\sigma _{\mu_3} &\simeq& 0.19\,R^3\,\sigma_m \label{theor_3rd_sim} \\
\sigma _{v_{\rm R}} &\simeq& 0.43\, R\,\omega\,\sigma_m \label{theor_vR_sim}
\end{eqnarray}
Comparing these relations with the theoretical results in (\ref{theor_pos})--(\ref{theor_vr}) we find that the numerical
factors from the numerical experiments are systematically some 30--40\% larger than according to the statistical theory. This
discrepancy largely vanishes if the models are constrained to high inclinations ($i\simeq\pm 90^\circ$). This suggests that
the discrepancy is mainly caused by the small values of $\sigma_m$ obtained in models with small inclinations, i.e.\ when the
star is seen nearly pole-on. The differences in these factors are in any case well within the scatter indicated in
Eq.~\ref{theor_pos_sim}--\ref{theor_vR_sim}, which emphasizes the statistical nature of the predictions based e.g.\ on
photometric variations.

It should also be noted that there is a considerable scatter
between the different realisations reported in
Eqs.~(\ref{res1})--(\ref{res4}),
amounting to about 50\% RMS about the mean RMS effect. Thus, any
prediction based on either (\ref{theor_pos})--(\ref{theor_vr}) or
(\ref{res1})--(\ref{res4}) is only valid in a statistical sense,
with considerable uncertainty in any individual case. Nevertheless,
the overall agreement between the results of these very different
models suggests that the statistical relations among the different
effects have a fairly general validity. The expressions
for $\sigma_{v_{\rm R}}$ are the least general in this respect, as
they obviously break down if the structures change on a time scale
smaller than $P$, or if the surface structures themselves have
velocity fields. Equations (\ref{theor_pos}) and (\ref{theor_3rd})
do not depend on the assumption that the variability is caused by
the rotation.

When modeling spotted stars, any brightening effect of faculae is often disregarded \citep[for more details
see][]{Ulvas2005A&A...435.1063A}; only the darkening effect of spots is computed. For the Sun, the effect of faculae is known
to be comparable and sometimes even larger than the darkening effect of sunspots
\citep{Eker2003A&A...404.1107E,Chapman1984Natur.308..252C,Chapman1986SoPh..103...21C,%
Chapman1992JGR....97.8211C,Steinegger1996ApJ...461..478S}. However, since
the general relationships, e.g.\ in (\ref{theor_pos})--(\ref{theor_vr}),
are equally valid for bright and dark spots (or any mixture of them),
it should still be possible to predict the astrometric effects from
the photometric variations.

\subsection{Comparison with previous studies and observations}

The (near-) proportionality between the observable effects and the spot filling factor $f\propto AN$ expressed by
Eqs.~(\ref{Saar 1})--(\ref{Hatzes 2}) is not supported by our spotted model, which predicts that the effects are proportional
to $A\sqrt{N}$. However, for small $N$ and a filling factor of a few percent we have rough quantitative agreement with these
earlier results. We note that (\ref{Hatzes 1}) and (\ref{Hatzes 2}) can be combined to give an approximate relation similar
to (\ref{vr_vsini}).

Equation (\ref{H.-Gs relation}) derived by \citet{Ludwig2005} is
practically identical to our (\ref{theor_pos}), which is not
surprising as they are based on very similar statistical models.

Both the theoretical result and the result from the simulation for
the relationship between the RMS for the radial velocity
and the RMS for the magnitude shows a distinct relation and this
result is confirmed by observations in the literature
\citep{Paulson_Saar2004AJ....127.1644P} for a very limited number
of stars in the Hyades all having rotation period of $P\sim 8.5$~days.
These are G0V--G5V stars and should therefore have approximately the
same radii as the Sun ($R\sim 7\times 10^5$~km).
Equation~(\ref{theor_vR_sim}) then gives
\begin{eqnarray}
\sigma _{v_{\rm R}} \simeq 2600\,\sigma_m\,\,[\mbox{m~s}^{-1}]
\end{eqnarray}
in reasonable agreement with the empirical result in (\ref{Paulson fig4}).
The simulations by \citet{sozzetti2005} give an astrometric jitter that is
roughly a factor 2 greater than predicted by (\ref{theor_pos}) or
(\ref{theor_pos_sim}).

Thus the results of previous studies generally agree within a factor 2 or
better with the theoretical formulae derived in this Section.

\section{Application to real stars}

In this section we use known statistics about the photometric and radial-velocity
variations of real stars in order to predict the expected astrometric jitter for
different types of stars. Rather than using angular units, we consistently
express the astrometric jitter in linear units, using the astronomical unit AU,
mAU ($10^{-3}$~AU) or $\mu$AU ($10^{-6}$~AU). This eliminates the dependence
on the distance to the star, while providing simple conversion to angular units:
$1~\mu$AU corresponds to $1~\mu$as at a distance of 1~pc. We also note
that $1~\mbox{mAU}\simeq 0.215R_\odot$ and $1~\mu\mbox{AU}\simeq 150$~km.

\subsection{Pre-Main Sequence (T Tauri) stars}

T Tauri stars are low-mass, pre-main sequence stars in a dynamic stage of
evolution often characterised by prominent dark spots, bipolar outflows or
jets, accreting matter with associated rapid brightness variations, and
in many cases circumstellar disks extending to a few hundred AU
\citep[e.g.,][]%
{Rhode2001AJ....122.3258R,Herbst2002A&A...396..513H,Aurora2005AJ....129..363S}.
Taking the star-forming region in the Orion nebula as an example, the spectral
types range from G6 to M6, with the large majority in the range K0 to M4
\citep{Rhode2001AJ....122.3258R}.

Many processes may contribute to the astrometric jitter of these stars besides their surface structures, e.g.\ photometric
irregularities of the circumstellar disk. The statistical relations derived in Sect.~\ref{theory} could therefore mainly set
a \emph{lower} limit to the likely astrometric effects. \cite{Herbst1994AJ....108.1906H} found that the photometric
variability of (weak) T Tauri stars (WTTS) is of the order of 0.8~mag due to cool spots and occasional flares. Assuming a
typical radius of $\sim 2R_{\odot}$ \citep{Rhode2001AJ....122.3258R}, Eq.~(\ref{theor_pos_sim}) leads to an estimated
astrometric variability of the order of $1R_{\odot }\sim 5000~\mu$AU.

\subsection{Main-Sequence stars}

\citet{Eyer_Grenon1997hipp.conf..467E} have used the Hipparcos photometric data to
map the intrinsic variability of stars across the HR diagram. On the main sequence
(luminosity class V), stars of spectral type B8--A5 and F1--F8 are among the most
stable ones, with a mean intrinsic variability $\sigma_m<2$~mmag and with only a
few percent of the stars having amplitudes above $0.05$~mag. Early B type stars are
nearly all variable with a mean intrinsic variability of $\sim 10$~mmag, and
among the cool stars the level and frequency of variability increases from late G
to early M dwarfs. In the instability strip (A6--F0) the main-sequence stars are
mostly micro-variable with $\sigma_m$ up to several mmag.
Among F--K stars the degree of variability is probably also a strong function
of age or chromospheric activity \citep{Fekel2004}; e.g., the Hyades
(age $\sim 600$~Myr) show variations of about 10~mmag \citep{Radick1995}.

The Sun (G2V) is located in one of the photometrically most stable parts of
the main sequence, and is one of the (as yet) few stars for which the micro-variability
has been studied in detail. Analysis of the VIRGO/SoHO total solar irradiance data
\citep{Lanza2003} show variability at the level $\sigma_m\simeq 0.25$~mmag
(relative variance $5\times 10^{-8}$) on time scales $\lesssim 30$~days, which
can largely be attributed to rotational modulation. The longer-term,
solar-cycle related variations are of a similar magnitude. The optical data
show a strong wavelength dependence, with $\sigma_m\simeq 0.2$~mmag at 860~nm
increasing to 0.4~mmag at 550~nm and 0.5~mmag at 400~nm \citep{Lanza2004}.
For comparison, a single large sunspot group (equivalent area
$A=0.05\%$, corresponding to $f=0.1\%$) gives $\sigma_m\simeq 0.6$~mmag
according to (\ref{res1}).

The photometric variations of the Sun on short (rotation-related) timescales appears to be representative for solar-like
stars of similar age and chromospheric activity \citep{Fekel2004}. Thus, we may expect $\sigma_m\lesssim 1$~mmag for `solar
twins' candidates, such as the sample studied by \citet{Melendez2006ApJ...641L.133M}. Inspection of the Hipparcos photometry
for these stars \citep{ESA1997} confirm that most of them show no sign of variability at the sensitivity limit of a few mmag.
Much more detailed and accurate statistics on micro-variability in solar-type stars are soon to be expected as a result of
survey missions such as MOST \citep{Walker2003}, COROT \citep{Baglin2002} and Kepler \citep{Basri2005}.

The increased frequency and amplitude of variations for late G-type and cooler dwarf stars is at least partly attributable to
starspots. \citet{Aigrain2004} estimated stellar micro-variability as function of age and colour index from a scaling of the
solar irradiance power spectrum based on the predicted chromospheric activity level. For example, they find $\sigma_m\simeq
1.5$~mmag in white light for old ($\sim 4.5$~Gyr) F5--K5 stars, practically independent of spectral type, while for young
stars ($\sim 625$~Myr) $\sigma_m$ increases from 2 to 7~mmag in the same spectral range.

Variability among field M dwarfs has been studied e.g.\ by
\citet{Rockenfeller2006A&A...448.1111R}, who find that a third of the stars
in their sample of M2--M9 dwarfs are variable at the level of
$\sigma_m\sim 20$~mmag. Evidence for large spots has been found for many
K and M stars, yielding brightness amplitudes of up to a few tenths of a
magnitude.

A large body of data on radial-velocity jitter in (mainly) F, G and K stars has been assembled from the several on-going
planet search programmes and can be used to make statistical predictions as function of colour, chromospheric activity and
evolutionary stage. However, since at least part of the radial-velocity jitter is caused by other effects than the rotation
of an inhomogeneous surface (e.g., by atmospheric convective motions), its interpretation in terms of astrometric jitter is
not straight-forward. From the observations of $\sim$450 stars in the California and Carnegie Planet Search Program,
\citet{Wright2005} finds a radial velocity jitter of $\sim\,$4~m~s$^{-1}$ for inactive dwarf stars of spectral type F5 or
later, increasing to some 10~m~s$^{-1}$ for stars that are either active or more evolved. \citet{Saar+Butler1998}, using data
from the Lick planetary survey, find intrinsic radial-velocity jitters of 2--100~m~s$^{-1}$ depending mainly on rotational
velocity ($V\sin i$) and colour, with a minimum around $B\!-\!V\simeq 1.0$--1.3 (spectral type $\sim$K5). For a sample of
Hyades F5 to M2 dwarf stars, \citet{Paulson2004} find an average rms radial velocity jitter of $\sim$16~m~s$^{-1}$.

\subsection{Giant stars}\label{sec:giants}

For giants of luminosity class III, Hipparcos photometry has shown a considerable
range in the typical degree of variability depending on the spectral type
\citep{Eyer_Grenon1997hipp.conf..467E}.
The most stable giants ($\sigma_m<2$~mmag) are the early A and late G types.
The most unstable ones are of type K8 or later, with a steadily increasing
variability up to $\sim\,$0.1~mag for late M giants.
The stars in the instability strip (roughly from A8 to F6) are typically
variable at the 5--20~mmag level. As these are presumably mainly radially
pulsating, the expected astrometric jitter is not necessarily higher than
on either side of the instability strip. This general picture is confirmed by
other studies. \citet{jorissen+97} found that late G and early K giants are
stable at the $\sigma_m\le 6$~mmag level; K3 and later types have an
increasing level of micro-variability with a time scale of 5 to 10~days,
while $b-y=1.1$ ($\simeq$M2) marks the onset of large-amplitude
variability ($\sigma_m\ge 10$~mmag) typically on longer time scales
($\sim 100$~days). From a larger and somewhat more sensitive survey of
G and K giants, \citet{henry2000} found the smallest fraction of
variables in the G6--K1 range, although even here some 20\% show
micro-variability at the 2--5~mmag level; giants later than K4 are
all variable, half of them with $\sigma_m\ge 10$~mmag. The onset of
large-amplitude variability coincides with the coronal dividing line
\citep{haisch+91} separating the earlier giants with a hot corona
from the later types with cool stellar winds. This suggests that the
variability mechanisms may be different on either side of the dividing
line, with rotational modulation of active regions producing the
micro-variability seen in many giants earlier than K3 and pulsation
being the main mechanism for the larger-amplitude variations in the later
spectral types \citep{henry2000}.

Several radial-velocity surveys of giants
\citep{Frink2001,Setiawan2004,Hekker2006} show increasing intrinsic
radial-velocity variability with $B\!-\!V=1.2$, with a more or less
abrupt change around $B\!-\!V=1.2$ ($\simeq\,$K3). Most bluer giants
have $\sigma_{v_{\rm R}}\simeq 20$~m~s$^{-1}$ while the redder ones
often have variations of 40--100~m~s$^{-1}$.

\subsection{Bright giants and supergiants}\label{sec:supergiants}

With increasing luminosity, variability becomes increasingly common among the bright giants and supergiants (luminosity class
II--Ia). The Hipparcos survey \citep{Eyer_Grenon1997hipp.conf..467E} shows a typical intrinsic scatter of at least 10~mmag at
most spectral types, and of course much more in the instability strip (including the cepheids) and among the red supergiants
(including semiregular and irregular variables). Nevertheless there may be a few `islands' in the upper part of the
observational HR diagram where stable stars are to be found, in particular around G8II.

It is clear that pulsation is a dominating variability mechanism for many of these objects. However, `hotspots' and other
deviations from circular symmetry has been observed in interferometrical images of the surfaces of M supergiants and Mira
varibles \citep[e.g.,][]{tuthill+97,tuthill+99}, possibly being the visible manifestations of very large convection cells,
pulsation-induced shock waves, patchy circumstellar extinction, or some other mechanism. Whatever the explanation for these
asymmetries may be, it is likely to produce both photometric and astrometric variations, probably on time scales of months to
years. \citet{Kiss2006} find evidence of a strong $1/f$ noise component in the power spectra of nearly all red supergiant
semiregular and irregular variable stars in their sample, consistent with the picture of irregular variability caused by
large convection cells analogous to the granulation-induced variability background seen for the Sun.

\subsection{Summary of expected astrometric jitter}

Table~\ref{table:1} summarises much of the data discussed in this Section
for the main-sequence, giant and supergiant stars, and gives the corresponding
estimates of the astrometric jitter ($\sigma_{\rm pos}$) based on theoretical
formulae. These estimates are given in three columns labelled with the
corresponding equation number:
\begin{itemize}
\item
Equation~(\ref{theor_pos}) is used to predict the positional jitter from
the typical values of photometric variability in column $\sigma_m$.
This is based on the assumption that the variability is due either to
(dark or bright) spots, granulation, or any other surface features that
vary with time. Note that the temporal variation need not be related to
stellar rotation. The resulting $\sigma_{\rm pos}$ are probably realistic
order-of-magnitude estimates except when the photometric variability is
mainly caused by radial pulsations. In such cases (e.g., for stars in the
instability strip and red supergiants) the values given clearly represent
upper limits to the real effect.
\item
Equation~(\ref{theor_vr}) is used to predict the astrometric effect from
the radial-velocity variability in column $\sigma_{v_{\rm R}}$. This
is only valid if the radial velocity is rotationally modulated.
Since pulsations, non-radial oscillations, convection and many other
effects may cause radial-velocity variations without a corresponding
astrometric effect, these estimates are again upper limits. Nevertheless,
rotational modulation is important among active (young) main-sequence
stars and M dwarfs, and for these objects Eq.~(\ref{theor_vr}) may provide
correct order-of-magnitude estimates.
\item
Finally we have included an estimate of the astrometric jitter based on
the following equation
\begin{equation}\label{eq:sl}
\sigma_{\rm pos}=(300~\mu\mbox{AU})\times 10^{1-\log g}
\end{equation}
with $\log g$ taken from \citet{AQ}. Equation~(\ref{eq:sl}) is derived from the inverse relation to surface gravity $g$ found
by \citet{Svensson_Ludwig_2005} for a range of hydrodynamical model atmospheres. Although the authors warn that sphericity
effects may render an extrapolation of this relation to supergiants very uncertain, we have applied it to all the stellar
types in the table. Since it only includes the random effects of stellar granulation, it represents a lower limit to the
expected astrometric jitter.
\end{itemize}
If the estimates based on the photometric and radial-velocity estimates are strictly considered as upper limits, the results
in the table appear rather inconclusive. However, if the likely mechanisms of the variabilities are also considered, it is
possible to make some quantitative conclusions. For main-sequence A to M stars, the expected level of astrometric jitter is
generally in the range 2--20~$\mu$AU probably depending mainly on the level of stellar activity; old, inactive stars should
have less jitter (2--5~$\mu$AU). The Sun appears to be more stable than the typical old, solar-like star, but not by a large
factor. The most stable giant stars are the late F to early K types, were the expected astrometric jitter is of order
25~$\mu$AU. Late-type giants and supergiants have $\sigma_{\rm pos}$ of a hundred to several thousand $\mu$AU.

\begin{table*}[tbp]
\caption{A summary of typical photometric and spectroscopic variability
for different stellar types, and inferred levels of astrometric jitter
($\sigma_{\rm pos}$). The jitter is estimated in three different ways:
from the photometric variability, using Eq.~(\ref{theor_pos}) [this will
overestimate the jitter if part of the variability is due to radial pulsation];
from the radial-velocity variability, using Eq.~(\ref{theor_vr}) [this method
will overestimate the jitter if the variability is not caused by rotational
modulation]; and from the surface gravity, using Eq.~(\ref{eq:sl}) [this
only includes jitter caused by granulation, and is therefore a lower
limit]. References to typical observed quantities are given as footnotes.
Radii and $\log g$ (not shown) are taken from \citet{AQ}.}
\label{table:1}
\centering
\begin{tabular}{llccccccc}
\hline\noalign{\smallskip}
\multicolumn{2}{l}{Type} & $\sigma_m$ & $\sigma_{v_{\rm R}}$ & $R$ & $P$
& $\sigma_{\rm pos}$ (\ref{theor_pos}) & $\sigma_{\rm pos}$ (\ref{theor_vr})
& $\sigma_{\rm pos}$ (\ref{eq:sl}) \\
&     & [mmag]     & [m~s$^{-1}$]       & [$R_\odot$] & [d] & [$\mu$AU] & [$\mu$AU] & [$\mu$AU]\\
\noalign{\smallskip}\hline\noalign{\smallskip}
\multicolumn{6}{l}{Main sequence stars:} \\
&O--B7V  & 10$^c$    && 7 && 120 && 0.3\\
&B8--A5V & $<$2$^c$  && 2.5 && $<$9 && 0.2\\
&A6--F0V & 2--8$^c$  && 1.6 && 5--20 && 0.1\\
&F1--F8V & $<$2$^c$  & 3--100$^m$ & 1.3 & 3$^b$ & $<$5 & 1--30 & 0.1\\
&F9--K5V (young) & 5--15$^{a,d,k}$ & 16$^j$ & 1 & 10$^a$ & 10--25 & 18 & 0.1\\
&F9--K5V (old) & 1--3$^{a,d}$ & 3--5$^{k}$ & 1 & 25$^a$ & 2--5 & 8--14 & 0.1\\
&G2V (Sun) & 0.4$^i$ & & 1 & 25$^b$ & 0.7 && 0.1\\
&K6--M1V & 10$^c$ & 5$^m$ & 0.6 & 40$^a$ & 10 & 20 & 0.1\\
&M2--M9V & 20$^l$ & 10$^m$ & 0.3 & 0.2--2$^l$ & 10 & 0.2--2 & 0.04\\
\noalign{\smallskip}
\multicolumn{6}{l}{Giants:} \\
&O--B7III & 4--8$^c$  && 10 && 70--140 && 1\\
&B8--A7III & $<$4$^c$ && 5 && $<$35 && 1.5\\
&A8--F6III & 5--20$^c$ && 5 && 50--200 && 2\\
&F7--G5III & 2--6$^c$ & $<$20$^f$ & 7 & 10$^b$ & 25--75 & $<$25 & 5\\
&G6--K2III & $<$2$^{c,g}$ & 20--30$^{e,f,n}$ & 15 & 30$^b$ & $<$50 & 60 & 20\\
&K3--K8III & 5--10$^{c,h}$ & 20--100$^{e,f,n}$ & 25 && 200--500 && 50\\
&M0III & 20$^{c,h}$ & 30--150$^{e,f,n}$ & 40 && 1400 && 150\\
&M5III & 100$^{c,h}$ & 50--300$^{e,f,n}$ & 90 && 16000\\
\noalign{\smallskip}
\multicolumn{6}{l}{Bright giants and supergiants:} \\
&O--AIa,b  & 4--40$^c$  && 30 && 200--2000 && 25\\
&FIa,b & 20--100$^{d}$ && 100 && 4000--20\,000 && 100\\
&GII & 2--10$^c$  && 30 && 100--500 && 40\\
&G--KIa,b & 10--100$^c$ && 150 && 3000--30\,000 && 250\\
&MIa,b,II & $\sim$100$^c$ && 500 && $\sim$100\,000 && 300--3000\\
\noalign{\smallskip}\hline \noalign{\smallskip\scriptsize
References:
$^a$\cite{Aigrain2004},
$^b$\cite{AQ},
$^c$\cite{Eyer_Grenon1997hipp.conf..467E},
$^d$\cite{Fekel2004},
$^e$\cite{Frink2001},
$^f$\cite{Hekker2006},
$^g$\cite{Henry2002ApJ...577L.111H},
$^h$\cite{jorissen+97},
$^i$\cite{Lanza2004},
$^j$\cite{Paulson_Saar2004AJ....127.1644P},
$^k$\cite{Radick1995},
$^l$\cite{Rockenfeller2006A&A...448.1111R},
$^m$\cite{Saar+Butler1998},
$^n$\cite{Setiawan2004}   }

\end{tabular}
\end{table*}

\section{Discussion}

\subsection{Astrometric signature of exoplanets}\label{sec:exoplanets}

The possibility for an astrometric detection of a planet depends on the angular size of the star's wobble on the sky relative
to the total noise of the measurements, including the astrophysically induced astrometric jitter discussed in the previous
section. In linear measure, the size of the wobble is approximately given by the semi-major axis of the star's motion about
the common centre of mass, or the \emph{astrometric signature}
\begin{equation}\label{eq:alpha}
\alpha=\frac{M_{\rm p}}{M_\ast+M_{\rm p}}\, a
\simeq\frac{M_{\rm p}}{M_\ast}\, a
\end{equation}
\citep[cf.][who however express this as an angle]{Lattanzi+2000},
where $M_{\rm p}$ is the mass of the exoplanet, $M_\ast$ that of the star,
and $a$ the semi-major axis of the relative orbit. In all cases of interest
here, $M_{\rm p}\ll M_\ast$, so that the second equality can be used.

It is of interest to evaluate the astrometric signature for the already
detected exoplanets. For most of them we only know $M_{\rm p}\sin i$ from
the radial-velocity curve, and we use this as a proxy for $M_{\rm p}$.
This somewhat underestimates the astrometric effect, but not by a large
factor since the spectroscopic detection method is strongly biased against
systems with small $\sin i$. Analysing the current (April 2007) data in
the Extrasolar Planets Encyclopaedia \citep{Schneider_exoplanets} we find
a median value $\alpha\simeq 1200~\mu$AU; the 10th and 90th percentiles
are 15 and 10\,000~$\mu$AU.

Future exoplanet searches using high-precision astrometric techniques may however primarily target planets with masses in the
range from 1 to 10~Earth masses ($M_{\rm Earth} \simeq 3\times 10^{-6}M_\odot$) in the habitable zone of reasonably
long-lived main-sequence stars (spectral type A5 and later, lifetime $\la 1$~Gyr). For a star of luminosity $L$ we may take
the mean distance of the habitable zone to be $a\sim (L/L_\odot)^{1/2}$~AU
\citep{Kasting1993Habitable_zones,Gould2003ApJ...591L.155G}. In this mass range ($\sim 0.2$--$2~M_\odot$) the luminosity
scales as $M_\ast^{4.5}$ \citep[based on data from][]{Andersen91}, so we find $a\propto M_\ast^{2.25}$ and
\begin{equation}\label{eq:alpha}
\alpha \simeq \left(3~\mu\mbox{AU}\right)\times
\left(\frac{M_{\rm p}}{M_{\rm Earth}}\right)
\left(\frac{M_\ast}{M_\odot}\right)^{1.25}
\end{equation}
For a planet of one Earth mass orbiting a main-sequence star, this
quantity ranges from about 7~$\mu$AU for an A5V star to 2.3~$\mu$AU
for spectral type K0V.

\citet{Lopez+2005} have argued that life will have time to develop also
in the environments of subgiant and giant stars, during their slow phases
of development. The habitable zone may extend out to 22~AU for a
1~$M_\odot$ star, with a correspondingly larger astrometric signature.
However, the long period of such planets would make their detection
difficult for other reasons.


\subsection{Exoplanet detection}
\label{sec:disce}

The detection probability is in reality a complicated function of many factors such as the number of observations, their
temporal distribution, the period and eccentricity of the orbit, and the adopted detection threshold (or probability of false
detection). A very simplistic assumption might be that detection is only possible if the RMS perturbation from the planet
exceeds the RMS noise from other causes. Neglecting orbital eccentricity and assuming that the orbital plane is randomly
oriented in space, so that $\langle\sin^2 i\rangle=2/3$, the RMS positional excursion of the star in a given direction on the
sky is $\alpha/\sqrt{3}$. With a sufficiently powerful instrument, so that other error sources can be neglected, the
condition for detection then becomes $\alpha/\sigma_{\rm pos}\ga \sqrt{3}$. In reality, a somewhat larger ratio than
$\sqrt{3}$ is probably required for a reliable detection, especially if the period is unknown. For example,
\citet{sozzetti2005} reports numerical simulations showing that $\alpha/\sigma\ga 2$ is required for detection of planetary
signatures by SIM or Gaia, where $\sigma$ is the single-epoch measurement error, provided that the orbital period is less
than the mission length. (For the corresponding problem of detecting a periodic signal in radial-velocity data,
\citet{Marcy+2005} note that a velocity precision of 3~m~s$^{-1}$ limits the detected velocity semi-amplitudes to greater
than $\sim$10~m~s$^{-1}$, implying an even higher amplitude/noise ratio of 3.3.) As a rule-of-thumb, we assume that detection
by the astrometric method is at least in principle possible if
\begin{equation}\label{eq:det}
\sigma_{\rm pos}\la 0.5\alpha
\end{equation}
For old, solar-type stars the expected astrometric jitter is
$\la$5~$\mu$AU, implying that exoplanets around these stars with
$\alpha \ga 10~\mu$AU could generally be detected and measured
astrometrically. This applies to the vast majority ($>$90\%) of the
exoplanets already detected by the radial-velocity method. Such
observations would be highly interesting for obtaining independent
information about these systems, in particular orbital inclinations
and unambiguous determination of planetary masses.

Exoplanets of about 10~$M_{\rm Earth}$ orbiting old F--K main-sequence stars in the habitable zone ($\alpha\simeq
20$--50~$\mu$AU) would generally be astrometrically detectable. This would also be the case for Earth-sized planets in
similar environments ($\alpha\simeq 2$--5~$\mu$AU), but only around stars that are unusually stable, such as the Sun.

\subsection{Determination of parallax and proper motion}

The primary objective of high-precision astrometric measurements, apart from
exoplanet detection, is the determination of stellar parallax and proper
motion. We consider here only briefly the possible effects of stellar surface
structures on the determination of these quantities.

Stellar parallax causes an apparent motion of the star, known as the parallax
ellipse, which is an inverted image the Earth's orbit as viewed from the star.
The linear amplitude of the parallax effect is therefore very close to 1~AU.
(For a space observatory at the Sun--Earth Lagrangian point L$_2$, such as
Gaia, the mean amplitude is 1.01~AU.) Thus, the size of the astrometric jitter
expressed in AU can directly be used to estimate the minimum achievable
\emph{relative} error in parallax. For main-sequence stars this relative error
is less than $10^{-4}$, for giant stars it is of order $10^{-4}$ to $10^{-3}$,
and for supergiants it may in some cases exceed 1\%. We note that a 1\% relative
error in parallax gives a 2\% (0.02~mag) error in luminosity or absolute
magnitude.

If proper motions are calculated from positional data separated by $T$ years,
the random error caused by the astrometric jitter, converted to transverse velocity,
is $\simeq\sigma_{\rm pos}\sqrt{2}/T$. Even for a very short temporal baseline
such as $T=1$~yr, this error is usually very small: $\sim$0.1~m~s$^{-1}$ for
main-sequence stars and $\sim$0.5--5~m~s$^{-1}$ for giants.
(Note that $1~\mbox{AU~yr}^{-1}\simeq 4.74~\mbox{km~s}^{-1}$.) In most
applications of stellar proper motions this is completely negligible.

\section{Conclusions}

For most instruments on ground or in space, stars are still unresolved or marginally resolved objects that can only be
observed by their disk-integrated properties. The total flux, astrometric position, effective radial velocity and closure
phase are examples of such integrated properties. Stellar surface structures influence all of them in different ways. Our
main conclusions are:
\begin{enumerate}
\item
Theoretical considerations allow to establish statistical relations between the different integrated properties of stars.
Under certain assumptions these relations can be used to predict the astrometric jitter from observed variations in
photometry, radial velocity or closure phase.
\item
The total flux, astrometric position and third central moments (related to closure phase) are simple moments of the intensity
distribution over the disk, and for these the statistical relations are valid under fairly general conditions -- for example,
they hold irrespective of whether the variations are caused by spots on a rotating star or by the temporal evolution of
granulation. By contrast, radial-velocity variations can only be coupled to photometric and astrometric variations if they
are primarily caused by rotational modulation.
\item
The theoretical relations are supported by numerical simulations using a model of a rotating spotted star. In this case the
variations in total flux, position, radial velocity and closure phase are all proportional to $A\sqrt{N}$, where $A$ is the
equivalent area of each spot and $N$ the number of spots. This means that, e.g., the astrometric jitter can be
(statistically) predicted from the photometric variability without knowing $A$ and $N$. It is noted that the spot filling
factor, being proportional to $AN$, is not the most relevant characteristic of spottiness for these effects.
\item
Using typical values for the observed photometric and radial-velocity variations in ordinary stars, we have estimated the
expected size of the astrometric jitter caused by surface structures (Table~\ref{table:1}). The estimates range from below
1~$\mu$AU for the Sun, several $\mu$AU for most main-sequence stars, some tens of $\mu$AU for giants, and up to several mAU
for some supergiants.
\item
The expected positional jitter has implications for the possible astrometric detection of exoplanets. While planets heavier
than 10~Earth masses may be astrometrically detected in the habitable zone around ordinary main-sequence stars, it is likely
that Earth-sized planets can only be detected around stars that are unusually stable for their type, similar to our Sun.
\item
Stellar surface structures in general have negligible impact on other astrometric applications, such as the determination of
parallax and proper motion. A possible exception are supergiants, where very large and slowly varying spots or convection
cells could limit the relative accuracy of parallax determinations to a few per cent.
\end{enumerate}

\begin{acknowledgements}
We give special thanks to Dainis Dravins, Jonas Persson and Andreas Redfors for helpful discussions and comments on the
manuscript, and to Hans-G\"{u}nter Ludwig for communicating his results from simulations of closure phase. We also thank
Kristianstad University for funding the research of UE and thereby making this work possible.
\end{acknowledgements}

\bibliographystyle{aa}
\bibliography{spot_ref2}

\appendix

\section{Statistical properties of the spatial moments of the
intensity distribution across a stellar disk} \label{appendixA}

In this Appendix we derive the mean values and variances of the moments $\langle x^m y^n \rangle$ for a spherical star, where
$x$ and $y$ are spatial coordinates normal to the line-of-sight and $\langle\rangle$ denotes the instantaneous flux-weighted
mean. The analysis extends and generalizes that of \citet{Ludwig2005} by considering also the third moment (relevant for
measurement of closure phase) and a centre-to-limb variation of the intensity contrast.

Let $\theta,\varphi$ be polar coordinates on the stellar surface with $\theta=0$ at the centre of the visible disk and
$\varphi=0$ along the $x$ axis. With $\mu=\cos\theta$ we write the instantaneous intensity across the visible stellar surface
$S$ as $I(\mu,\varphi)$, and introduce the non-normalized spatial moments
\begin{eqnarray}\label{a1}
\lefteqn{M_{mn}\equiv\int_S\mbox{d}S\,I(\mu,\varphi)\,\mu\,x^m\,y^n=}\nonumber\\
&& R^{m+n+2}\!\!\int_0^1 \!\!\mbox{d}\mu\! \int_{0}^{2\pi }\!\!\!\mbox{d}\varphi
\,I(\mu,\varphi)\,\mu\,(1-\mu^2)^{(m+n)/2}\cos^m\!\varphi\,\sin^n\!\varphi
\end{eqnarray}
where $x=R\,(1-\mu^2)^{1/2}\cos\varphi$ and $y=R\,(1-\mu^2)^{1/2}\sin\varphi$.
The factor $\mu$ in the integrand is the foreshortening of the surface
element $\mbox{d}S=R^2\mbox{d}\mu\,\mbox{d}\varphi$ when projected normal
to the line-of-sight. The normalized moments are given by
\begin{equation}\label{a2}
\langle x^m y^n \rangle = \frac{M_{mn}}{M_{00}}
\end{equation}
where it can be noted that $M_{00}$ equals the instantaneous total stellar flux.

It is assumed that $I(\mu,\varphi)$ varies randomly both across the stellar
surface (at a given instant), and as a function of time. As a consequence,
the spatial moments (\ref{a1}) and (\ref{a2}) are also random functions of
time, and the goal is to characterize them in terms of their mean values
and variances.\footnote{We use the notation $\mbox{E}[X]$ for the mean value
(expectation) of the generic random variable $X$,
$\mbox{V}[X]=\mbox{E}[(X-\mbox{E}[X])^2]$ for the variance and
$\mbox{D}[X]=\mbox{V}[X]^{1/2}$ for the rms dispersion.}
Since we are interested in quite small effects of the surface structure it is
generally true that the dispersions are small compared with the total flux and
scale of the star, so that for example $\mbox{D}[M_{00}]\ll\mbox{E}[M_{00}]$
and $\mbox{D}[M_{10}]\ll R\,\mbox{E}[M_{00}]$. In this case the variability
of $\langle x^m y^n \rangle$ is mainly produced by the numerator in (\ref{a2}),
and we may use the approximations
\begin{equation}\label{a3}
\mbox{E}[\langle x^m y^n \rangle] = \frac{\mbox{E}[M_{mn}]}{\mbox{E}[M_{00}]}\, ,
\quad
\mbox{D}[\langle x^m y^n \rangle] = \frac{\mbox{D}[M_{mn}]}{\mbox{E}[M_{00}]}
\end{equation}
In the following we therefore focus on deriving the mean values and dispersions
of the non-normalized moments $M_{mn}$.
The (temporal) mean value and dispersion of $I(\mu,\varphi)$ are assumed to be
independent of $\varphi$; thus
\begin{equation}\label{a4}
\mbox{E}[I(\mu,\varphi)] = A(\mu)\, ,
\quad
\mbox{D}[I(\mu,\varphi)] = \sigma_I(\mu)
\end{equation}
where $A(\mu)$ and $\sigma_I(\mu)$ are functions to be specified.

\subsection{Mean value of the moments}

We assume a linear limb-darkening law with coefficient $a$, such that
\begin{equation}\label{a5}
A(\mu) = (1-a+a\mu)A_1
\end{equation}
where $A_1$ is the mean intensity at the disk centre ($\mu=1$). From (\ref{a1})
we obtain
\begin{eqnarray}\label{a6}
\lefteqn{\mbox{E}[M_{mn}]=}\nonumber\\
&& R^{m+n+2}\!\!\int_0^1 \!\!\mbox{d}\mu\! \int_{0}^{2\pi }\!\!\!\mbox{d}\varphi
\,A(\mu)\,\mu\,(1-\mu^2)^{(m+n)/2}\cos^m\!\varphi\,\sin^n\!\varphi
\end{eqnarray}
which with (\ref{a5}) evaluates to
\begin{equation}\label{a7}
\mbox{E}[M_{mn}]= 0
\end{equation}
if either $m$ or $n$ is odd, and to
\begin{equation}\label{a8}
\mbox{E}[M_{mn}]= 2\pi R^{m+n+2}\,\frac{(m-1)!!(n-1)!!}{(m+n)!!}\,A_1 H_{m+n}(a)
\end{equation}
if both $m$ and $n$ are even.%
\footnote{The double factorial notation means $k!!=k(k-2)(k-4)\cdots 2$
for even integer $k$, and $k!!=k(k-2)(k-4)\cdots 1$ for odd $k$. We have
$0!!=(-1)!!=1$.}
Here, we introduced the functions
\begin{eqnarray}\label{a9}
H_k(a) &=& \int_0^1 \!\!\mbox{d}\mu\,(1-a+a\mu)\,\mu\,(1-\mu^2)^{k/2}\nonumber\\
&=& \left\{ \begin{array}{lll}
{\displaystyle\frac{1-a}{k+2} + \frac{k!!}{(k+3)!!}\,a} && \mbox{(even~$k$)}\\[12pt]
{\displaystyle\frac{1-a}{k+2} + \frac{k!!}{(k+3)!!}\,\frac{\pi}{2}\,a} && \mbox{(odd~$k$)}
\end{array}\right.
\end{eqnarray}
of which, presently, we only need
\begin{equation}\label{a10}
H_0(a)=\frac{1}{2}-\frac{a}{6} \qquad \mbox{and} \qquad
H_2(a)=\frac{1}{4}-\frac{7a}{60}
\end{equation}
Thus, the mean total flux is
\begin{equation}\label{a11}
\mbox{E}[M_{00}]= \pi R^2\,A_1\,\left( 1 - \frac{a}{3} \right)
\end{equation}
and the second moments
\begin{equation}\label{a12}
\mbox{E}[M_{20}] = \mbox{E}[M_{02}] = \frac{1}{4}\,\pi R^4\,A_1\,\left( 1 - \frac{7a}{15} \right)
\end{equation}
The rms extension of the star in either coordinate is given by
\begin{equation}\label{a13}
s = \left( \frac{\mbox{E}[M_{20}]}{\mbox{E}[M_{00}]}\right)^{1/2}
= \left( \frac{\mbox{E}[M_{02}]}{\mbox{E}[M_{00}]}\right)^{1/2} = \frac{R}{2}\,
\left( \frac{1-7a/15}{1-a/3} \right)^{1/2}
\end{equation}

\subsection{Dispersion of the moments}

In order to compute the dispersion of $M_{mn}$ we need to introduce the
second-order statistics of $I(\mu,\varphi)$. Following \citet{Ludwig2005}
we divide the visible hemisphere into $N$ equal surface patches of size
$\Delta A =R^2\Delta\theta\Delta\varphi\sin\theta = R^2\Delta\mu\Delta\varphi
= 2\pi R^2/N$, with the centre of patch $k$ at position $(\mu_k,\varphi_k)$.
Thus the integral over the visible surface of any function $g(\mu,\varphi)$
can in the limit of large $N$ be replaced by a sum:
\begin{equation}\label{a14}
R^2\!\!\int_0^1 \!\!\mbox{d}\mu\! \int_{0}^{2\pi }\!\!\!\mbox{d}\varphi
\,g(\mu,\varphi) \simeq \frac{2\pi R^2}{N} \sum_{k=1}^N g(\mu_k,\varphi_k)
\end{equation}
In particular, from (\ref{a1}) we have
\begin{equation}\label{a15}
M_{mn} = \frac{2\pi R^{m+n+2}}{N}\sum_{k=1}^N
I_k\,\mu_k\,(1-\mu_k^2)^{(m+n)/2}\cos^m\!\varphi_k\,\sin^n\!\varphi_k
\end{equation}
where $I_k$ is the mean value of $I(\mu,\varphi)$ in patch $k$.
This expresses the moment as a linear combination of the random variables
$I_k$. If we now assume that the intensity variations
$\Delta I_k=I_k-\mbox{E}[I_k]$ of the patches are uncorrelated,
i.e.\ $\mbox{E}[\Delta I_k\Delta I_{k'}]=0$ for $k\ne k'$, we have
\begin{eqnarray}\label{a16}
\lefteqn{\mbox{V}[M_{mn}] =}\nonumber\\
&& \frac{4\pi^2 R^{2m+2n+4}}{N^2}\sum_{k=1}^N
\mbox{V}[I_k]\,\mu_k^2\,(1-\mu_k^2)^{m+n}\cos^{2m}\!\varphi_k\,\sin^{2n}\!\varphi_k
\end{eqnarray}
For sufficiently large $N$ the patches would resolve even the smallest surface
structures and we would have $\mbox{V}[I_k]=\sigma_I^2(\mu_k)$ according to
(\ref{a4}). However, in that case the intensities of adjacent patches would be
correlated, so (\ref{a16}) would not hold. For the latter equation we effectively
need patches that are larger than the correlation length of the surface structures.
We must therefore assume that $N$ is large enough for the discretization (\ref{a12})
to hold, and still small enough that the patches are uncorrelated. In this regime
we have $\mbox{V}[I_k]<\sigma_I^2(\mu_k)$, since $I_k$ is the \emph{average} intensity
in patch $k$, not the \emph{local} intensity at point $(\mu_k,\varphi_k)$. In fact,
$\mbox{V}[I_k]$ will depend on the patch size (or $N$) in such a way that
$\mbox{V}[I_k]/N$ is invariant \citep{Ludwig2005}. (This is obviously the case for
independent patches: grouping them into larger and fewer patches decreases the
variance in proportion to the resulting $N$.) We write the invariant quantity as
\begin{equation}\label{a17}
\frac{\mbox{V}[I_k]}{N} = A(\mu_k)^2 C(\mu_k)^2
\end{equation}
where $A(\mu)$ is the mean intensity as before and $C(\mu)$ the centre-to-limb
variation of the contrast (scaled by $N^{-1/2}$). In analogy with (\ref{a5}) we
assume a linear centre-to-limb variation of the contrast according to
\begin{equation}\label{a18}
C(\mu) = (1-c+c\mu)C_1
\end{equation}
Inserting (\ref{a17}) and (\ref{a18}) into (\ref{a16}) and using (\ref{a14}) to
transform the sum into an integral gives
\begin{eqnarray}\label{a19}
\lefteqn{\mbox{V}[M_{mn}] = 2\pi R^{2m+2n+4} \times}\nonumber\\
&& \times\int_0^1 \!\!\mbox{d}\mu\! \int_{0}^{2\pi }\!\!\!\mbox{d}\varphi
\,A(\mu)^2 C(\mu)^2\,\mu^2\,(1-\mu^2)^{m+n}\cos^{2m}\!\varphi\,\sin^{2n}\!\varphi \nonumber\\
&=& 4\pi^2 R^{2m+2n+4}\,\frac{(2m-1)!!(2n-1)!!}{(2m+2n)!!}\,A_1^2 C_1^2\, K_{m+n}(a,c)
\end{eqnarray}
where we introduced the functions
\begin{equation}\label{a20}
K_k(a,c) = \int_0^1 \!\!\mbox{d}\mu\,(1-a+a\mu)^2(1-c+c\mu)^2\mu^2(1-\mu^2)^k
\end{equation}
For $k=0,\dots 3$ we have
\begin{eqnarray}\label{a21}
K_0(a,c) &=& \frac{1}{3} - \frac{1}{6}\,(a+c) + \frac{1}{30}\,(a^2+4ac+c^2)\nonumber\\
&&\qquad - \frac{1}{30}\,ac(a+c) + \frac{1}{105}\,a^2c^2\\
K_1(a,c) &=& \frac{2}{15} - \frac{1}{10}\,(a+c) + \frac{1}{42}\,(a^2+4ac+c^2)\nonumber\\
&&\qquad- \frac{11}{420}\,ac(a+c) + \frac{1}{126}\,a^2c^2\\
K_2(a,c) &=& \frac{8}{105} - \frac{29}{420}\,(a+c) + \frac{23}{1260}\,(a^2+4ac+c^2)\nonumber\\
&&\qquad - \frac{3}{140}\,ac(a+c) + \frac{47}{6930}\,a^2c^2\\
K_3(a,c) &=& \frac{16}{315} - \frac{13}{252}\,(a+c) + \frac{29}{1980}\,(a^2+4ac+c^2)\nonumber\\
&&\qquad - \frac{25}{1386}\,ac(a+c) + \frac{38}{6435}\,a^2c^2
\end{eqnarray}
Using (\ref{a3}) we obtain the following general expression for the dispersion of the
normalized spatial moment $D_{mn}\equiv \mbox{D}[\langle x^m y^n\rangle ]$:
\begin{equation}\label{a22}
D_{mn} = C_1\,R^{m+n}\,\frac{\sqrt{\displaystyle
\frac{(2m-1)!!^{\phantom{'}}(2n-1)!!}{(2m+2n)!!}\,K_{m+n}(a,c)}}{H_0(a)}
\end{equation}
Note that $D_{00}=\mbox{D}[M_{00}]/\mbox{E}[M_{00}]$ is the relative dispersion of the
total flux, $D_{10}$ is the dispersion of the photocentre along the $x$ axis, etc.
We have in particular
\begin{eqnarray}\label{a23}
D_{00} &=& C_1\,\frac{\sqrt{\!^{\phantom{'}}K_0(a,c)}}{H_0(a)}\\
D_{10} &=& D_{01} = C_1\,R\,\frac{\sqrt{\frac{1}{2}K_1(a,c)}}{H_0(a)}\\
D_{20} &=& D_{02} = C_1\,R^2\,\frac{\sqrt{\frac{3}{8}K_2(a,c)}}{H_0(a)}\\
D_{30} &=& D_{03} = C_1\,R^3\,\frac{\sqrt{\frac{5}{16}K_3(a,c)}}{H_0(a)}
\end{eqnarray}

\subsection{The third central moment}

Closure phase is sensitive to the asymmetry of the stellar image, and the
third moments ($M_{mn}$ for $m+n=3$) are intended to provide a statistical
characterization of this asymmetry. These moments are calculated with respect
to the \emph{geometrical centre} of the disk (at $x=y=0$). However, intrinsic
image properties such as size, shape and asymmetry are more properly expressed
with respect to the \emph{photocentre}, at $x_0=M_{10}/M_{00}$,
$y_0=M_{01}/M_{00}$, i.e., by means of central moments (here denoted with
a prime). For example, the third central moment along the $x$ axis is given by
\begin{eqnarray}\label{a24}
\lefteqn{M_{30}'=\int_S\mbox{d}S\,I(\mu,\varphi)\,\mu\,(x-x_0)^3}\nonumber\\
&& \quad = M_{30} - 3x_0 M_{20} + 3x_0^2 M_{10} - x_0^3 M_{00}\nonumber\\
&& \quad = M_{30} - 3\,\frac{M_{10}M_{20}}{M_{00}} + 2\,\frac{M_{10}^3}{M_{00}^2}
\end{eqnarray}
It is seen that $\mbox{E}[M_{30}']=0$ as expected. However, to calculate
the variance of $M_{30}'$ it is necessary to make some approximations. First we
replace the even moments in the right-hand side of (\ref{a23}) by their mean
values and introduce the rms extent $s$ of the stellar disk from (\ref{a13}),
yielding
\begin{equation}\label{a25}
M_{30}' = M_{30} - 3M_{10}\left(s^2 - \frac{2}{3}\, x_0^2\right)
\end{equation}
The photocentre displacement is normally very small compared with the size of
the disk, so that the second term in the parentheses can be neglected. Then
\begin{eqnarray}\label{a26}
\lefteqn{\mbox{V}[M_{30}']=\mbox{E}[M_{30}'^2]
= \mbox{E}[M_{30}^2] - 6s^2\mbox{E}[M_{10}M_{30}] + 9s^4\mbox{E}[M_{10}^2]}\nonumber\\
&& \quad = \mbox{V}[M_{30}] - 6s^2\mbox{V}[M_{20}] + 9s^4\mbox{V}[M_{10}]
\end{eqnarray}
using that $\mbox{E}[M_{10}M_{30}]=\mbox{V}[M_{20}]$. Finally, the dispersion
of the normalized third central moment is found to be
\begin{eqnarray}\label{a27}
D_{30}' &=& \frac{C_1 R^3}{H_0(a)}\left\lgroup\displaystyle
\frac{5}{16}\,K_3(a,c)-\frac{9}{8}\,\frac{H_2(a)}{H_0(a)}\,K_2(a,c)+
\phantom{\left(\frac{H_2(a)}{H_0(a)}\right)^2}\right.\nonumber\\
&&\left.\qquad\qquad{}+\frac{9}{8}\,\left(\frac{H_2(a)}{H_0(a)}\right)^2
K_1(a,c)\right\rgroup^{1/2}
\end{eqnarray}

\subsection{Scaling relations}

\begin{figure}[hbt]
  \resizebox{88mm}{!}{%
  \includegraphics[angle=-90]{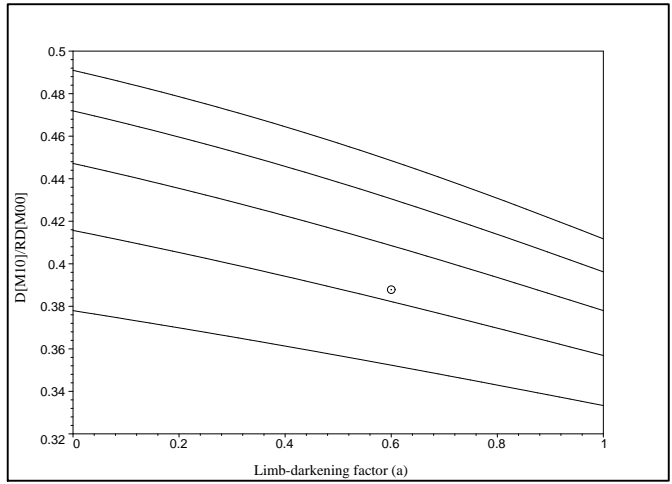}}
  \caption{The scaling factor $\mbox{D}[M_{10}]/R\mbox{D}[M_{00}]$ from Eq.~(\ref{a28})
  between the expected dispersions in photocentre position and total stellar flux,
  plotted as function of the limb-darkening parameter $a$ and the parameter for the
  centre-to-limb variation of surface structure contrast ($c$). The different curves represent, from top to bottom,
  $c$=-1, -0.5, 0, 0.5, 1. The solar symbol indicates
  the typical value for solar granulation in white light, $a=0.6$ and $c=0.4$.
}
  \label{figA1}
\end{figure}

\begin{figure}[hbt]
  \resizebox{88mm}{!}{%
  \includegraphics[angle=-90]{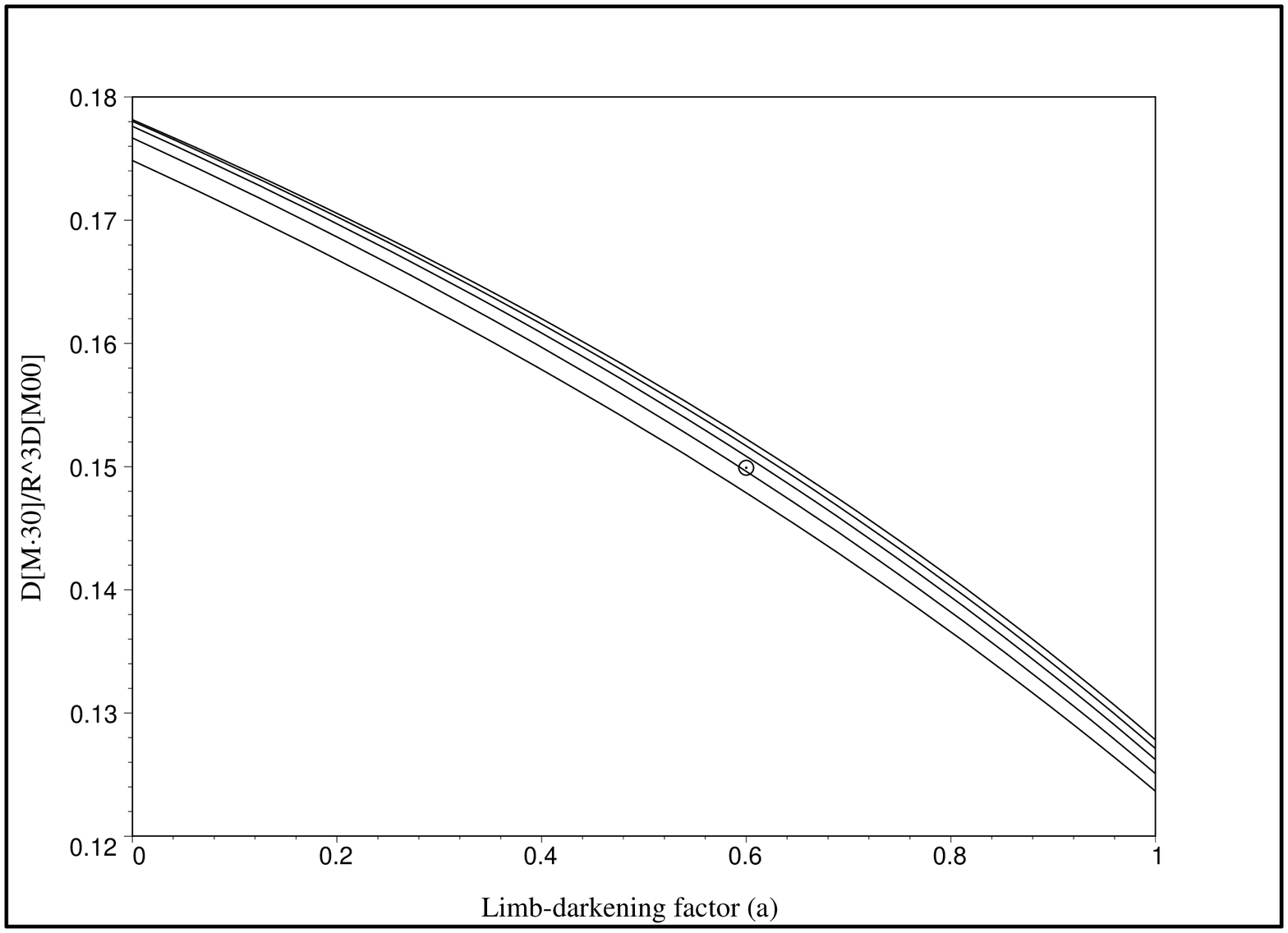}}
  \caption{Similar to \figurename~\ref{figA1} but for the scaling factor $\mbox{D}[M_{30}']/R^3\mbox{D}[M_{00}]$ from Eq.~(\ref{a29})}
  \label{figA2}
\end{figure}

Since all the dispersions in (\ref{a23}) and (\ref{a27}) are proportional to
$C_1$ we obtain a set of simple scaling relations that can be used to predict
the dispersion of a certain moment from a measurement of the dispersion of
a different moment (assuming that $a$ and $c$ are approximately known).
The most useful relations allow to predict the dispersions of the
first and third moments from that of the total flux (zeroth moment).
For the first moment (photocentre) we have
\begin{equation}\label{a28}
\frac{\mbox{D}[M_{10}]}{\mbox{D}[M_{00}]}
= R\left\lgroup\frac{K_1(a,c)}{2K_0(a,c)}\right\rgroup^{1/2}
\end{equation}
and for the third central moment
\begin{eqnarray}\label{a29}
\lefteqn{\frac{\mbox{D}[M_{30}']}{\mbox{D}[M_{00}]}
= R^3\left\lgroup\displaystyle
\frac{5}{16}\,\frac{K_3(a,c)}{K_0(a,c)}-\frac{9}{8}\,\frac{H_2(a)}{H_0(a)}\,
\frac{K_2(a,c)}{K_0(a,c)}+
\phantom{\left(\frac{H_2(a)}{H_0(a)}\right)^2}\right.}\nonumber\\
&&{}\left.\qquad\qquad\qquad{}+\frac{9}{8}\,\left(\frac{H_2(a)}{H_0(a)}\right)^2 \frac{K_1(a,c)}{K_0(a,c)}\right\rgroup^{1/2}
\end{eqnarray}
The numerical factors on the right-hand sides of (\ref{a28}) and (\ref{a29})
are graphically shown in Figs.~\ref{figA1}--\ref{figA2} as functions of the
structural parameters $a$ and $c$.

\end{document}